\documentclass[12pt]{article}
\usepackage{multicol}
\usepackage{amssymb}
\usepackage{amsfonts,amssymb,amsmath,amsthm}
\usepackage{epsfig}
\usepackage{slashed}
\usepackage{graphicx}
\usepackage{subfig}
\usepackage{mathrsfs}
\usepackage{bbm}
\usepackage{cite}
\usepackage{enumerate}
\usepackage{slashbox}
\usepackage{color}
\usepackage[colorlinks,linkcolor=red,anchorcolor=red,citecolor=green]{hyperref}
\makeatletter 
\@addtoreset{equation}{section}
\makeatother  

\makeatletter

\newcommand{\Rmnum}[1]{\expandafter\@slowromancap\romannumeral #1@}
\makeatother

\begin{document}
\renewcommand{\thefootnote}{\fnsymbol{footnote}}
\begin{titlepage}

\vspace{10mm}
\begin{center}
{\Large\bf Microscopic structures and thermal stability of black holes conformally coupled to scalar fields in five dimensions}
\vspace{16mm}

{{\large Yan-Gang Miao${}^{}$\footnote{\em E-mail: miaoyg@nankai.edu.cn}
and Zhen-Ming Xu}${}^{}$\footnote{\em E-mail: xuzhenm@mail.nankai.edu.cn}

\vspace{6mm}
${}^{}${\normalsize \em School of Physics, Nankai University, Tianjin 300071, China}

}

\end{center}

\vspace{10mm}
\centerline{{\bf{Abstract}}}
\vspace{6mm}
\noindent
Completely from the thermodynamic point of view, we explore the microscopic character of a hairy black hole of Einstein's  theory conformally coupled to a scalar field in five dimensions by means of the Ruppeiner thermodynamic geometry. We demonstrate that the scalar hairy black hole has rich microscopic structures in different parameter spaces. Moreover, we analyze the thermal stability of this black hole in detail.
\vskip 20pt
\noindent
{\bf PACS Number(s)}: 04.50.Gh, 04.60.-m, 04.70.-s, 05.70.Ce


\end{titlepage}

\renewcommand{\thefootnote}{\arabic{footnote}}
\setcounter{footnote}{0}
\setcounter{page}{2}
\pagenumbering{arabic}

\section{Introduction}
It is known that a black hole, most likely as a bridge to connect general relativity and quantum mechanics, is a fascinating object due to its peculiarities uncovered recently. The pioneering works by Hawking and Bekenstein~\cite{SH,JDB} disclose that the black hole can possess temperature and entropy. Thereafter, this mysterious gravity system is mapped to a thermodynamic system, making the black hole thermodynamics receive a wide range of attention and also acquire a great deal of progress~\cite{JMB,SHD,RW,CEJM,DSJT,BPD,KM}. Even so, in comparison with an ordinary thermodynamic system, the exploration of the microscopic character of black holes is still a great challenge. The reason is that we do not have a complete quantum theory of gravity yet. The most likely candidate theories about quantum gravity are string theory and loop quantum gravity theory. They probably provide a natural framework of microscopic origins of black holes and explain most of the microscopic properties with the help of gravity/gauge duality and gravity/CFT correspondence~\cite{STR1,STR2,STR3,CR}. Intuitively, the Ruppeiner thermodynamic geometry might offer some guidance about the microscopic character of black holes completely from the thermodynamic viewpoint and the relevant studies in recent years indeed support~\cite{BMMZ,NTW,WL,DSM,ZDSM} such a viewpoint. Hence in the present work we are interested in the purely thermodynamic method and use it to explore the microstructure of black holes phenomenologically.

The Ruppeiner geometry~\cite{GR,GR2} is a thermodynamic geometry that is established on the language of Riemannian geometry, whose initial applications are focused on investigating an ordinary thermodynamic system. More precisely, this particular geometrical method is based on the theory of fluctuations of equilibrium thermodynamics and on the assumption that the Hessian is positive definite~\cite{GR3,ALT}. Now the Ruppeiner geometry is dealt with as a new attempt to extract the microscopic interaction information from the axioms of thermodynamics. The geometrical method has been applied successfully in fluid, solid and discrete lattice systems, see, for instance, some review articles~\cite{GR2,GR3}. In particular, due to the fact that black holes can be treated as a typical thermodynamic system, one can acquire some insights into microscopic structures of black holes by using the Ruppeiner thermodynamic method, although the constituents of black holes are still unclear.

In this paper, we will focus on microscopic structures and thermal stability of a specific hairy black hole of Einstein's theory with a scalar field conformally coupled to higher-order Euler densities in $D=5$ dimensions~\cite{OR,GLOR,GGO,GGGO,HM}. The reason is that this model has the following novel advantages:
\begin{itemize}
  \item There is much interest in the higher-curvature corrections to general relativity in the context of the AdS/CFT correspondence. The Lovelock theory has been proved to be a favorite playground to perform such corrections. But it has some limitations. For instance, when investigating $R^3$ terms in the 5-dimensional Lovelock theory which is also dubbed a quasi-topological gravity~\cite{Oliva,MR,MPS}, one demands higher-derivative couplings or non-minimal couplings with additional matter fields. In other words, although it provides a tractable model of higher-curvature gravity, the Lovelock theory should admit an extension of the scalar field matter conformally coupled to gravity in a non-minimal way and permit simple black hole solutions that present a back-reaction of the matter in arbitrary dimensions. As expected, the theoretical model we are currently interested in, as the generalization of the Lovelock theory, can offer a practical way with matter conformally coupled to gravity and with the back-reaction of the scalar field, i.e. a scalar field conformally coupled to higher-order Euler densities~\cite{MC,MOGJ}. This model as a kind of quasi-topological gravity can be regarded as the most general scalar field/gravity coupling formulation whose field equations are of second order for both gravity and matter. In this sense, it is also a generalization of Horndeski scalar field/gravity coupling theory~\cite{GWH} which is the most general real scalar field theory in $D=4$ dimensions with second order field equations.
  \item With a vanishing cosmological constant, the scalar hairy black holes of general relativity do exist in the 4-dimensional spacetime. For some explicit solutions, the scalar field configuration diverges on the horizon~\cite{JDB1,BBM,JDB2}, but for other solutions, the scalar field does not diverge~\cite{CF}. While for $D>4$ dimensions, the hairy black hole solutions do not exist~\cite{BTD}. With a nonvanishing cosmological constant, the corresponding black hole solutions were just found in $D=3, 4$ dimensions~\cite{CMJZ,MTZ,AAHM,BM1} or in some gauged supergravity for asymptotically AdS spacetimes~\cite{FKN}. However, for the special model we are currently discussing, which consists of scalar matter conformally coupled to gravity through a non-minimal coupling between a real scalar field and the dimensionally extended Euler densities (see details in Section 3), it does admit hairy black hole solutions in asymptotically flat and asymptotically (A)dS spaces in arbitrary $D$ dimensions. So this special model evades the restriction of the well-known no-hair theorem.\footnote{The no-hair theorem states that the black hole solutions of the Einstein-Maxwell equations in general relativity can be completely characterized by only three externally observable classical parameters: mass, charge and angular momentum~\cite{Gravitation,Wald}.} That is to say, the scalar hair can exist besides the mass, charge and angular momentum hairs. Moreover, the configuration of scalar fields for the hairy black hole is regular everywhere outside and on the horizon.
  \item In addition, the scalar hairy black hole has special thermodynamic behaviors and critical phenomena in various regions of the parameter space of the black hole solution. They exhibit not only the van der Waals-type phase transition~\cite{HM,YZ}, but also a reentrant phase transition~\cite{HM} which usually occurs in higher curvature gravity theory.
\end{itemize}
Considering the above advantages of the hairy black hole, we investigate its thermal stability through the analysis of heat capacity. More importantly, we calculate the thermodynamic scalar curvature and reveal rich microscopic structures in different parameter spaces with the aid of the Ruppeiner thermodynamic geometry which can offer some important information about the microscopic character of black holes.

The paper is organized as follows. In section \ref{sec2}, we briefly review the Ruppeiner thermodynamic geometry. In section \ref{sec3}, we calculate the thermodynamic scalar curvature and the heat capacity so as to make an exploration of microscopic structures and thermal stability in detail for the five-dimensional scalar hairy black hole. Finally, we devote to drawing our conclusion in section \ref{sec4}.

\section{Ruppeiner thermodynamic geometry}\label{sec2}
The metric of the Ruppeiner thermodynamic geometry defined in the entropy representation plays the role of the thermodynamic potential~\cite{GR2,GR3},
\begin{equation}
g_{\alpha\beta}=-\frac{\partial^2 S}{\partial x^{\alpha}\partial x^{\beta}},
\end{equation}
where $x^{\alpha}$ represents thermodynamic quantities. However, for black hole thermodynamics one usually derives the expression of energy (or enthalpy $M$) of the black hole system, and thus gives the thermodynamic metric in the Weinhold energy form~\cite{FW,FW2},
\begin{equation}
g_{\alpha\beta}=\frac{1}{T}\frac{\partial^2 M}{\partial X^{\alpha}\partial X^{\beta}}, \label{rg}
\end{equation}
where $T$ is the Hawking temperature
and the coordinates $X^{\alpha}=(S,P,Q,...)$ usually denote thermodynamic quantities. Based on the metric eq.~(\ref{rg}), one can construct one thermodynamic invariant, i.e. the thermodynamic scalar curvature $R$ similar to that of general relativity. To this end, one first defines the Christoffel symbols by following the calculation in refs.~\cite{GR2,GR4},
\begin{equation}
\Gamma^{\alpha}_{\beta\gamma}=\frac12 g^{\mu\alpha}\left(\partial_{\gamma}g_{\mu\beta}+\partial_{\beta}g_{\mu\gamma}-\partial_{\mu}g_{\beta\gamma}\right), \label{rgc}
\end{equation}
and then writes the Riemannian curvature tensor,
\begin{equation}
{R^{\alpha}}_{\beta\gamma\delta}=\partial_{\delta}\Gamma^{\alpha}_{\beta\gamma}-\partial_{\gamma}\Gamma^{\alpha}_{\beta\delta}+
\Gamma^{\mu}_{\beta\gamma}\Gamma^{\alpha}_{\mu\delta}-\Gamma^{\mu}_{\beta\delta}\Gamma^{\alpha}_{\mu\gamma}. \label{rgr}
\end{equation}
Finally, one obtains the thermodynamic scalar curvature $R$,
\begin{equation}
R=g^{\mu\nu}{R^{\xi}}_{\mu\xi\nu}. \label{rgs}
\end{equation}

The scalar curvature $R$ is independent of the choice of the coordinates, which means that $R$ is the most fundamental measurement of a thermodynamic system. Physically,  a lot of studies have shown~\cite{WL,DSM,ZDSM,GR3,GR4} that the sign of $R$ offers a direct information about the character of the interaction among molecules in a thermodynamic system. Concretely, there are three kinds of situations: i) a positive $R$ implies a repulsive interaction among molecules, ii) a negative $R$ implies an attractive interaction, and iii) a vanishing $R$ implies no interaction. In other words, $R>0$ ($R<0$) mimics the ideal Fermi (Bose) gas, and $R=0$ the classical ideal gas~\cite{JM,NS}.

\section{Hairy black hole in five dimensions}\label{sec3}
In order for the present work to be self-contained, we review the theory that consists of scalar matter conformally coupled to gravity through a non-minimal coupling between a real scalar field and the dimensionally extended Euler densities~\cite{OR,GLOR,MC}. This theory can yield second order field equations for both gravity and matter. In $D$ dimensions, one needs to define a four-rank tensor in terms of the Riemann curvature tensor ${R_{\mu\nu}}^{\gamma\delta}$ and the derivatives of the scalar field $\phi$,
\begin{equation}
{S_{\mu\nu}}^{\gamma\delta}=\phi^2{R_{\mu\nu}}^{\gamma\delta}-2\delta_{[\mu}^{[\gamma}\delta_{\nu]}^{\delta]}\nabla_{\rho}\phi\nabla^{\rho}\phi
 -4\phi \delta_{[\mu}^{[\gamma}\nabla_{\nu]}\nabla^{\delta]}\phi+8\delta_{[\mu}^{[\gamma}\nabla_{\nu]}\phi \nabla^{\delta]}\phi,
\end{equation}
where $\delta^{\mu}_{\nu}$ is the Kronecker tensor.
Hence, one can write down the general action of the theory with the above four-rank tensor,
\begin{align}
I=&\int \text{d}^D x\sqrt{-g}\sum_{k=0}^{\left[\frac{D-1}{2}\right]}\frac{k!}{2^k}\delta^{\mu_1}_{[\alpha_1}\delta^{\nu_1}_{\beta_1}\cdots
\delta^{\mu_k}_{\alpha_k}\delta^{\nu_k}_{\beta_k]}\times \nonumber\\
&\left(a_k{R^{\alpha_1\beta_1}}_{\mu_1\nu_1}\cdots {R^{\alpha_k\beta_k}}_{\mu_k\nu_k}+b_k\phi^{D-4k}{S^{\alpha_1\beta_1}}_{\mu_1\nu_1}\cdots {S^{\alpha_k\beta_k}}_{\mu_k\nu_k}\right),
\end{align}
where $a_k$ and $b_k$ are arbitrary coupling constants, $[\frac{D-1}{2}]$ is the integer part of $\frac{D-1}{2}$, $g_{\mu\nu}$ is the metric with mostly plus signatures, and $g$ is its determinant, $g :=\text{det}(g_{\mu\nu})$. When all the couplings $b_k$ vanish, the theory reduces to the Lovelock theory and particularly, if $a_{k\neq 1}=0$, it reduces to general relativity. The field equations for gravity and the scalar field take the following forms,
\begin{equation}
G_{\mu\nu}=T_{\mu\nu},
\end{equation}
\begin{equation}
\sum_{k=0}^{\left[\frac{D-1}{2}\right]}\frac{(D-2k)b_k}{2^k}\phi^{D-4k-1}k!\delta^{\mu_1}_{[\alpha_1}\delta^{\nu_1}_{\beta_1}\cdots
\delta^{\mu_k}_{\alpha_k}\delta^{\nu_k}_{\beta_k]}{S^{\alpha_1\beta_1}}_{\mu_1\nu_1}\cdots {S^{\alpha_k\beta_k}}_{\mu_k\nu_k}=0,
\end{equation}
where
\begin{align*}
& G_{\mu}^{\nu}=-\sum_{k=0}^{\left[\frac{D-1}{2}\right]}\frac{a_k}{2^{k+1}}\delta^{\nu\lambda_1\cdots\lambda_{2k}}_{\mu\rho_1\cdots\rho_{2k}}
{R^{\rho_1\rho_2}}_{\lambda_1\lambda_2}\cdots{R^{\rho_{2k-1}\rho_{2k}}}_{\lambda_{2k-1}\lambda_{2k}}, \\
& T_{\mu}^{\nu}=\sum_{k=0}^{\left[\frac{D-1}{2}\right]}\frac{b_k}{2^{k+1}}\phi^{D-4k}\delta^{\nu\lambda_1\cdots\lambda_{2k}}_{\mu\rho_1\cdots\rho_{2k}}
{S^{\rho_1\rho_2}}_{\lambda_1\lambda_2}\cdots{S^{\rho_{2k-1}\rho_{2k}}}_{\lambda_{2k-1}\lambda_{2k}}.
\end{align*}

In order to illustrate the structure of this action, one considers the particular case $a_{k>1}=0$, and  writes the action as the following slightly simpler form,
\begin{equation}
I=\int \text{d}^D x\sqrt{-g}\left(a_0+a_1 R+\mathscr{L}_{m}(\phi, \nabla\phi)\right),  \label{action}
\end{equation}
where $a_0=-\Lambda/(8\pi)$, $a_1=1/(16\pi)$, $\Lambda$ is the cosmological constant, $R$ is the scalar curvature, and the matter Lagranian $\mathscr{L}_{m}(\phi, \nabla\phi)$ reads
\begin{equation*}
\mathscr{L}_{m}(\phi, \nabla\phi)=\sum_{k=0}^{\left[\frac{D-1}{2}\right]}b_k\phi^{D-4k}\frac{k!}{2^k}\delta^{\mu_1}_{[\alpha_1}\delta^{\nu_1}_{\beta_1}\cdots
\delta^{\mu_k}_{\alpha_k}\delta^{\nu_k}_{\beta_k]}{S^{\alpha_1\beta_1}}_{\mu_1\nu_1}\cdots {S^{\alpha_k\beta_k}}_{\mu_k\nu_k}.
\end{equation*}
Generally, the above action allows the existence of static spherically symmetric black hole solutions if the coupling matter terms satisfy certain relations in $D$ dimensions. The solution contains the metric,
\begin{equation}
\text{ds}^2=-f(r)dt^2+\frac{dr^2}{f(r)}+r^2 d{\Omega}^{2}_{D-2},  \label{msolu}
\end{equation}
and the scalar field configuration,
\begin{equation}
\phi=\frac{N}{r},  \label{ssolu}
\end{equation}
where $d\Omega_{D-2}^2$ is the square of line element on a $(D-2)$-dimensional unit sphere, $N$ is the dimensional constant, and the general form of function $f(r)$  takes the form,
\begin{equation}
f(r)=1-\frac{16\pi m}{(D-2)V_{\Omega}r^{D-3}}-\frac{16\pi Q}{(D-1)(D-2)r^{D-2}}-\frac{2\Lambda}{(D-1)(D-2)}r^2, \label{solu}
\end{equation}
where $V_{\Omega}$ is the volume of a $(D-2)$-dimensional unit sphere. In addition, $m$ is a constant of integration associated to the mass of black holes and $Q$ is denoted by
\begin{equation}
Q=\sum_{k=0}^{\left[\frac{D-1}{2}\right]}\frac{b_k(D-2k-1)(D-1)!}{(D-2k-1)!}N^{D-2k},
\end{equation}
and the dimensional constant $N$ and couplings $b_k$ must obey the following constraints to ensure the existence of the solution expressed by eqs.~(\ref{msolu}), (\ref{solu}), and (\ref{ssolu}),
\begin{align}
& \sum_{k=1}^{\left[\frac{D-1}{2}\right]}k\frac{b_k(D-1)!}{(D-2k-1)!}N^{2-2k}=0, \nonumber\\
& \sum_{k=0}^{\left[\frac{D-1}{2}\right]}[D(D-1)+4k^2]\frac{b_k(D-1)!}{(D-2k-1)!}N^{-2k}=0. \label{yueshu}
\end{align}

Now we restrict our consideration for a hairy black hole of Einstein's theory conformally coupled to a scalar field in five dimensions~\cite{GGO,GGGO,HM}. This model's metric can be written as follows,
\begin{equation}
\text{d}s^2=-f \text{d}t^2+f^{-1}\text{d}r^2+r^2 (\text{d}\theta_1^2+\sin^2 \theta_1\text{d}\theta_2^2+\sin^2 \theta_1\sin^2 \theta_2\text{d}\varphi^2),
\end{equation}
where $0\leq \theta_i<\pi$ $(i=1,2)$, $0\leq \varphi<2\pi$, and the lapse function $f$ takes the form,
\begin{equation}
f(r)=1-\frac{m}{r^2}-\frac{q}{r^3}+\frac{r^2}{l^2}.
\end{equation}
Here $l$ represents the effective AdS curvature radius that is associated with the cosmological constant $\Lambda$. Moreover, $q$ is characterized as the strength conformally coupled to a scalar field,
\begin{equation}
q=\frac{4\pi Q}{3}=\frac{64\pi}{5}b_1\left(-\frac{18b_1}{5b_0}\right)^{3/2}. \label{defq}
\end{equation}
Note that according to eq.~(\ref{yueshu}) there is an additional constraint: $10b_0 b_2=9b_1^2$, in order to ensure the existence of this black hole solution. 

\subsection{Thermodynamic analysis}
For the above mentioned hairy black hole in five dimensions,
its thermodynamic properties have been investigated in ref.~\cite{GGGO} and its reentrant phase transition and van der Waals behavior have been studied in ref.~\cite{HM}. In addition, we have discussed~\cite{YZ} the validity of Maxwell equal area law for this black hole in detail. Here we are just going to list out the thermodynamic quantities that we need in the following discussions. The thermodynamic enthalpy $M$, temperature $T$, and entropy $S$ take the following forms in terms of the event horizon,\footnote{The horizon satisfies the following equation,
$$f(r)=0 \qquad \Rightarrow \qquad r^5+l^2 r^3-ml^2r-ql^2=0.$$
Our analysis shows that the above equation must have real positive roots and the number of horizons depends on the relationship among the three parameters, $m,~q$ and $l$. In the present work we adopt the event horizon $r_h$, i.e. the largest real positive root of the above equation.}
\begin{eqnarray}
M&=&\frac{3\pi}{8}m=\frac{3\pi}{8}\left(r_h^2-\frac{q}{r_h}+\frac{r_h^4}{l^2}\right), \label{enth}\\ 
T&=&\frac{2r_h^3 l^2+ql^2+4r_h^5}{4\pi l^2 r_h^4}, \label{temp}\\
S&=&\frac{{\pi}^2}{2}\left(r_h^3-\frac{5}{2}q\right). \label{entr}
\end{eqnarray}
Moreover, the role of the cosmological constant $\Lambda$ is analogous to the thermodynamic pressure,
\begin{equation}
P=-\frac{\Lambda}{8\pi}=\frac{3}{4\pi l^2}, \label{pre}
\end{equation}
and the thermodynamic volume $V$ conjugate to the thermodynamic pressure $P$ has the form,
\begin{equation}
V \equiv \left(\frac{\partial M}{\partial P}\right)_{S,q}=\frac{\pi^2}{2}r_h^4.
\end{equation}

Now we analyze the thermodynamic stability of this black hole utilizing the heat capacity at constant pressure. The heat capacity can be calculated from eqs.~(\ref{enth}),~(\ref{temp}), and~(\ref{entr}), and expressed for a fixed $q$ in terms of the horizon $r_h$ as follows,
\begin{equation}
C_P\equiv T\left(\frac{\partial S}{\partial T}\right)_{P,q}=\frac{3\pi^2 r_h^3}{4}\cdot \frac{2r_h^3 l^2+q l^2+4r_h^5}{2r_h^5-r_h^3 l^2-2ql^2}.
\end{equation}
As was known, the black hole is locally stable for $C_P >0$, but locally unstable for $C_P <0$. In addition, a diverging heat capacity is a feature of second-order phase transitions in the ordinary fluid and spin systems, so as to black holes. The stability analyses that depend on the sign of the parameter $q$ can be summarized below.

\paragraph{Case $q>0$}
For this case, no $P-V$ critical phenomena\footnote{The $P-V$ critical phenomenon means that the behavior of a black hole is analogous to that of the van der Waals fluid when an oscillating part exists in the $P-V$ diagram.} occur~\cite{HM,YZ} and the extremal black hole that corresponds to the limit of zero temperature, $T\rightarrow 0$, does not exist. Nevertheless, for the heat capacity at constant pressure, there is only one divergence, which implies that the black hole will undergo one second-order phase transition from a locally unstable state to a locally stable one.

\paragraph{Case $q<0$}
In this situation, {according to eqs.~(\ref{temp}) and~(\ref{pre}), the equation of state for this specific black hole can be read as
\begin{equation}
P(r_h, T)=\frac{3T}{4r_h}-\frac{3}{8\pi r_h^2}-\frac{3q}{16\pi r_h^5}, \label{state1}
\end{equation}
and correspondingly the critical phenomenon occurs at the critical isotherm $T=T_c$ when the $P-V$ diagram has an inflection point,
\begin{equation}
\frac{\partial P}{\partial r_h}=0, \qquad \frac{\partial^2 P}{\partial r_h^2}=0. \label{creq}
\end{equation}
Solving the above equations, one can obtain the $P-V$ critical values~\cite{HM,YZ},
\begin{eqnarray}
r_c&=& (-5q)^{1/3}, \nonumber \\
T_c&=&-\frac{3}{20}\cdot \frac{(-5q)^{2/3}}{\pi q}, \nonumber\\
P_c&=&\frac{9}{200\pi}\left(-\frac{\sqrt{5}}{q}\right)^{2/3}. \label{c1}
\end{eqnarray}
In this case the extremal black hole (at zero temperature) has zero heat capacity. For a non-extremal black hole,
the behaviors of the heat capacity at constant pressure, $C_P$, can be summarized as the following three subcases:
\begin{itemize}
  \item When $P>P_c$, the heat capacity is positive, i.e. $C_P >0$, which implies that the black hole is in a stable state.
  \item When $P=P_c$, the heat capacity is divergent at the above critical point $r_c$, which means that the black hole will undergo one second-order phase transition from one locally stable state to another locally stable one. When $P=P_c$ but $r_h\neq r_c$, the heat capacity is positive, which implies a stable black hole.
  \item When $P<P_c$, the heat capacity diverges in two places, indicating that there exist two phase transitions for the black hole. The first phase transition happens from a locally stable phase to a locally unstable one at the small horizon radius, and the second occurs from a locally unstable phase to a locally stable one at the large horizon radius. When the pressure approaches the critical value, the two phase transitions will merge into one and the black hole is stable due to the positivity of the heat capacity.
\end{itemize}

\subsection{Thermodynamic curvature}
We set coordinates $X^{\alpha}=(S,P)$ for a fixed $q$, and calculate the thermodynamic scalar curvature $R$ for this hairy black hole and express the result in terms of the horizon radius $r_h$ (see Appendix for the details),
\begin{eqnarray}
R=-\frac{2l^2(4r_h^3+5q)}{3\pi^2 r_h^3(2r_h^3 l^2+q l^2+4r_h^5)}. \label{rads}
\end{eqnarray}

We can see that the above result reduces to that of the five-dimenional AdS black hole in the limit $q=0$,
\begin{eqnarray}
R\rightarrow -\frac{4l^2}{3\pi^2 r_h^3(l^2+2r_h^2)},
\end{eqnarray}
and then we deduce $R<0$, which implies that an attractive intermolecular interaction dominates in the five-dimenional AdS black hole system.

For the non-vanishing $q$, we have the following separate cases.

\paragraph{Case $q>0$}
According to eq.~(\ref{rads}), we know that the thermodynamic scalar curvature is negative, $R<0$, 
which implies that an attractive interaction dominates in the interior of this black hole. In other words, the microscopic structure inside the black hole is most likely analogous to that of the ideal Bose gas~\cite{JM}.

\paragraph{Case $q<0$}
In this situation, for the extremal black hole, i.e. at zero temperature, $T\rightarrow 0$, the thermodynamic scalar curvature $R$ is divergent;  for the non-extremal black hole, the sign of the thermodynamic scalar curvature $R$ only depends on the factor $(4r_h^3+5q)$. At $r_h=(-5q/4)^{1/3}=0.63r_c$, we obtain $R=0$, which indicates no intermolecular interactions in the black hole system. When  $r_h>0.63r_c$, we get $R<0$, which indicates that an attractive intermolecular interaction dominates in this black hole system. At last, when $r_h<0.63r_c$, we deduce $R>0$, which indicates that a repulsive intermolecular interaction dominates in this black hole system. Overall, for the $q<0$ situation, we can also say that the microscopic feature of this black hole is most likely similar to that of the ideal anyon gas~\cite{MM,MM2}.

So far we have presented the thermodynamic properties of the black hole in terms of the heat capacity and the thermodynamic scalar curvature of the system. Now we make a comment on these two descriptions that are independent of each other, i.e. the heat capacity is used for the analysis of stability, while the thermodynamic scalar curvature for the analysis of similarity of the hairy black hole to the Fermi, Bose or ideal gas.
In Ruppeiner thermodynamic geometry, the thermodynamic scalar curvature $R$ simply tells us the type of interaction, but no information of black hole constituents. For conventional statistical models, such as the Fermi, Bose or ideal gas system, the  constituents are known. Therefore, one can directly analyze stability and phase transitions. While for black holes, it is still unclear about their constituents, so that we adopt an opposite process to explore the constituents of black holes, i.e., from the type of interaction to microscopic structure. Hence, our statement is that the microscopic structure of the black hole is most likely analogous to that of the ideal Fermi, Bose or anyon gas. This conclusion is just a preliminary and phenomenological one which is derived from the similarity of the sign of thermodynamic curvature between the black hole and the ideal Fermi, Bose or anyon gas. This method might also be regarded as a new attempt to expand black hole thermodynamics.

\subsection{Thermodynamic curvature on the co-existence curve}

In this subsection, we investigate the microscopic structure in the small-large black hole\footnote{Here we regard the small black hole with the horizon radius $r_h$ below the critical value $r_c$ and the large black hole with the horizon radius $r_h$ above the critical value $r_c$.} co-existing phase. Our interest will focus on the $q<0$ situation with the $P-V$ parameters less than their critical values because only in this case does a rich microscopic structure exist. The co-existence curve of the two phases is determined by the Clausius-Clapeyron equation and the Maxwell equal area law. The Clausius-Clapeyron equation governs the slope of the co-existence curve of the small and large black hole phases. The Maxwell equal area law can remove an oscillating part in the $P-V$ diagram resulting that one can obtain the form of the small and large black hole phases. Note that the validity of Maxwell equal area law for this black hole has been studied in detail in our previous work~\cite{YZ}. The behaviors of the co-existence curve of two phases and thermodynamic curvature are shown in Figure \ref{tu}.
\begin{figure}
\begin{center}
  \begin{tabular}{cc}
    \includegraphics[width=68.5mm]{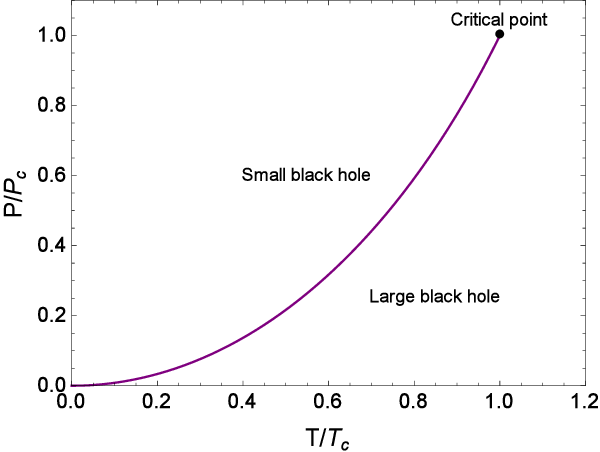} &
    \includegraphics[width=72mm]{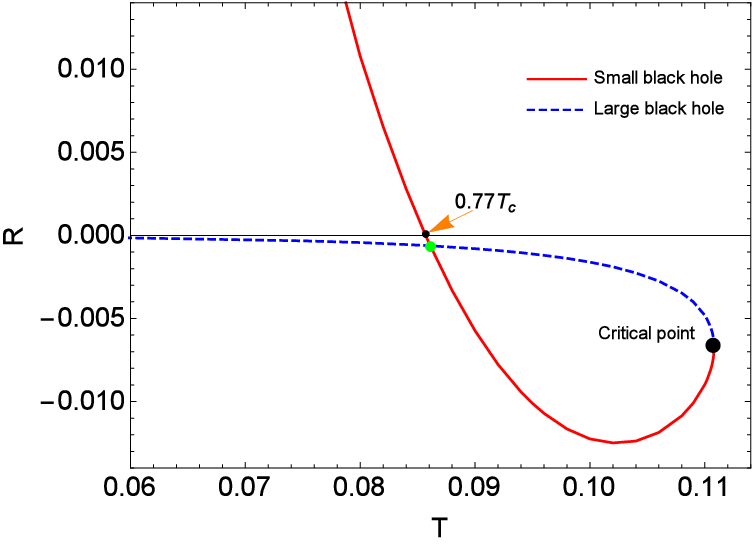} \\
  \end{tabular}
\end{center}
\caption{(color online) \textbf{Left}: The co-existence curve for the five-dimensional AdS hairy black hole. \textbf{Right}: The thermodynamic curvature $R$ with respect to temperature $T$ at $q = -2$ on the co-existence curve.}
\label{tu}
\end{figure}

From the left diagram of Figure \ref{tu}, we observe that the co-existence curve terminates at the critical point. The small and large black holes can clearly be distinguished from each other below the critical point, but they are mixed on the co-existence curve. However, we cannot distinguish one from the other above the critical point. From the right diagram of Figure \ref{tu}, we gain the following important information about the small and large black holes along the co-existence curve.
\begin{itemize}
  \item For the large black hole, the thermodynamic curvature is negative, $R<0$. Hence, we can say that the behaviors of this large black hole look much like the ideal Bose gas. With the decreasing of temperature $T$, the value of the thermodynamic curvature $R$ is almost close to zero, indicating the weakly-interacting constituents. As a result, the extremal large black hole resembles the ideal gas.
  \item For the small black hole, when $T=0.77T_c$, which corresponds to $r_h=0.63r_c$, the thermodynamic curvature is zero, $R=0$. When $T<0.77 T_c$, the thermodynamic curvature is positive, $R>0$, but when $0.77 T_c<T<T_c$, the thermodynamic curvature is negative, $R<0$. Hence, we observe that the non-extremal small black hole is most likely similar to the ideal anyon gas, while the extremal small black hole resembles the ideal Fermi gas.
  \item Both the small and large black holes share the same value of the thermodynamic curvature at two points. The first one is located at the critical point (black point) and the second at the green point in the right diagram of Figure \ref{tu}. The thermodynamic curvatures of the two points are negative, where the value of the critical point can be expressed as follows,
      \begin{eqnarray}
      R_c=\frac{2}{15\pi^2 q}.
      \end{eqnarray}
  \item All possible stable black holes possess the feature of low temperature $T$ and close zero thermodynamic curvature $R$. When the small-large black hole phase transition occurs, according to Figure 1, at temperature $T<0.77T_c$ we can see intuitively that it is the nature of repulsive intermolecular interaction on the small black hole side that produces a very strong outward degenerate pressure, causing the expansion of the small black hole  and leading to the formation of the large black hole, just as the prediction about the RN-AdS black hole in ref.~\cite{DSM}. While for temperature $0.77T_c<T<T_c$, because $R$ for the small black hole is negative and thus the interaction becomes attractive, and the interaction for the large black hole is also attractive, it is still unclear how to interpret the phase transition.
\end{itemize}

For the convenience of the following discussions, we introduce some reduced parameters defined by
\begin{eqnarray}
t:=\frac{T}{T_c}, \qquad \tilde{R}:=\frac{R}{|R_c|}, \qquad x_{s,l}:=\frac{r_{s,l}}{r_c},
\end{eqnarray}
where the subscript $s$ stands for the small black hole and $l$ for the large black hole. Note that $0<t\leq 1$, $x_s<1$ and $x_l >1$. So the thermodynamic curvature eq.~(\ref{rads}) can be written as
\begin{eqnarray}
\tilde{R}_{s,l}=\frac{1}{3t}\left(\frac{1}{x^7_{s,l}}-\frac{4}{x^4_{s,l}}\right).
\end{eqnarray}

Relevant studies have shown~\cite{YZ,XX} that when $t$ is decreasing, the reduced horizon radius of the small black hole $x_s$ decreases while that of the large black hole $x_l$ increases. Hence, for a very small $t$, we obtain the following asymptotic behaviors,
\begin{eqnarray}
\text{Small black hole:} \qquad \tilde{R}_s &\sim& \frac{1}{3t}\frac{1}{x^7_s},\\
\text{Large black hole:} \qquad \tilde{R}_l &\sim& -\frac{1}{3t}\frac{4}{x^4_l}.
\end{eqnarray}
From the above asymptotic formulations, we can conclude that for the small black hole, the positive thermodynamic curvature plays a dominating role, while for the large black hole, the negative thermodynamic curvature plays a dominating role. For the extremal small black hole, the thermodynamic curvature is positive (very large), but for the extremal large black hole, the thermodynamic curvature approaches to zero. Therefore, the asymptotic analyses are consistent with the previous numerical ones exhibited in Figure \ref{tu}.

\subsection{Special case without equivalent AdS background}
Reconsidering the primary action eq.~(\ref{action}), if set $a_0=0$, one can obtain~\cite{YZ2} a spherical symmetry solution without equivalent AdS background. Similar to our previous discussions, we can directly write the thermodynamic enthalpy $M$ and temperature $T$ in terms of the horizon $r_h$ as follows,
\begin{eqnarray}
M&=&\frac{3\pi}{8}\left(r_h^2-\frac{q}{r_h}\right),\\
T&=&\frac{2r_h^3+q}{4\pi r_h^4},
\end{eqnarray}
where the entropy $S$ takes the same expression as eq.~(\ref{entr}). Moreover, the heat capacity at a fixed $q$ can be read as
\begin{eqnarray}
C_q=-\frac{3\pi^2 r_h^3(2r_h^3+q)}{4(r_h^3+2q)},
\end{eqnarray}
and the thermodynamic curvature with respect to the coordinates $X^{\alpha}=(S,q)$ as
\begin{eqnarray}
R=\frac{4}{\pi^2 (2r_h^3+q)}.
\end{eqnarray}

We analyze the microscopic structure for the black hole system without equivalent AdS background and uncover its special features.

\paragraph{Case $q>0$}
In this situation, the heat capacity is negative, indicating that this black hole is in an unstable state. For the microscopic behavior, due to the positive thermodynamic curvature, $R>0$, we can clearly know that the black hole resembles the ideal Fermi gas and the repulsive intermolecular interaction dominates the black hole thermodynamic behaviors. 
At present, the Ruppeiner approach is regarded as a phenomenological description of the microstructure of black holes and it is independent of the usual thermodynamic one. By using only the Ruppeiner approach, we cannot temporarily analyze the stability of black holes, but it would be an interesting issue to establish a relationship between this approach and the stability.

\paragraph{Case $q<0$}
Under this circumstance, for the extremal black hole, i.e. $T=0$, the thermodynamic curvature $R$ is divergent. In addition, the thermodynamic curvature $R$ is positive for the non-extremal black hole ($T>0$). Hence, the behavior of this black hole is analogous to that of the ideal Fermi gas, where the repulsive intermolecular interaction dominates in this thermodynamic system. For the heat capacity $C_q$, there are four subcases that should be discussed in detail.
\begin{itemize}
\item For the extremal black hole, i.e. $T=0$, corresponding to $r_h=(-q/2)^{1/3}$, the heat capacity $C_q$ equals zero.
\item For the non-extremal black hole, i.e. $T>0$, the three different subcases are listed as follows:
\begin{itemize}
  \item When $(-q/2)^{1/3}<r_h<(-2q)^{1/3}$, the positive heat capacity $C_q>0$ implies that the black hole is locally stable.
  \item When $r_h=(-2q)^{1/3}$, the heat capacity is divergent, which means that the black holes will undergo one second-order phase transition from a locally stable state to a locally unstable one.
  \item When $r_h>(-2q)^{1/3}$, the negative heat capacity $C_q<0$ indicates that the black hole is locally unstable.
\end{itemize}
\end{itemize}
As a result, we obtain that for the black hole without equivalent AdS background in the cases of $q>0$ and $q<0$, the repulsive intermolecular interaction dominates in this thermodynamic system and its microscopic behavior matches that of the ideal Fermi gas. Moreover, this black hole is in an unstable state. Particularly, the black hole undergoes one second-order phase transition in the case of  $q<0$ and eventually it becomes more unstable.

\subsection{Treatment to the case with charges}
Here we only make a brief analysis. By adding the Maxwell action in five dimensions to eq.~(\ref{action}), we write the primary action~\cite{GGGO,HM},
\begin{equation}
I=\int \text{d}^5 x\sqrt{-g}\left(a_0+a_1 R-\frac{1}{16\pi}F^{\mu\nu}F_{\mu\nu}+\mathscr{L}_{m}(\phi, \nabla\phi)\right),
\end{equation}
where $F=dA$ with $A$ a 1-form potential. Correspondingly, the static spherically symmetric charged black hole solution can be directly written as
\begin{equation}
f(r)=1-\frac{m}{r^2}-\frac{q}{r^3}+\frac{e^2}{r^4}+\frac{r^2}{l^2},
\end{equation}
and the Maxwell potential is given by $A=\frac{\sqrt{3}e}{r^2}dt$, where $e$ is a constant of integration corresponding to the electric charge of the black hole.
Along the above way of investigations for that hairy black hole without charges and the way of discussions in refs.~\cite{HM,YZ}, we can directly calculate the heat capacity at constant pressure for a fixed coupling strength $q$ and a fixed charge $e$,
\begin{eqnarray}
C_P=\frac{3\pi^2 r_h^3(4r_h^6+2l^2 r_h^4+ql^2 r_h-2e^2 l^2)}{8r_h^6-4l^2(r_h^4+2q r_h)+20e^2 l^2},
\end{eqnarray}
and the thermodynamic curvature with coordinates $X^{\alpha}=(S,P)$ for fixed $q$ and $e$,
\begin{eqnarray}
R=-\frac{2l^2(4r_h^4+5q r_h-12e^2)}{3\pi^2 r_h^3(4r_h^6+2l^2 r_h^4+ql^2 r_h-2e^2 l^2)}.
\end{eqnarray}

Regardless of $q>0$ or $q<0$, the $P-V$ critical phenomena will occur and the extremal black hole will exist. For the extremal black hole, corresponding to $T=0$, the heat capacity at constant pressure is equal to zero and the thermodynamic curvature is divergent. For the non-extremal black hole, the heat capacity diverges at least  two times, indicating that the black hole undergoes two phase transitions. The first phase transition happens from a locally stable phase to a locally unstable one, and the second occurs from a locally unstable phase to a locally stable one. During the phase transitions, the sign of the thermodynamic curvature changes, undergoing the positive, negative and zero values. As a result, we can conclude that the microscopic feature of this black hole with charges is analogous to that of the ideal anyon gas~\cite{MM,MM2}.

\section{Summary}\label{sec4}
In refs.~\cite{GGGO,YZ}, some basic thermodynamic quantities, such as mass, entropy, temperature, and equation of state, have been studied. Meanwhile, the van der Waals-type behavior and the reentrant phase transitions have also been investigated in ref.~\cite{HM}. In the present work we gain more thermodynamic information of the hairy black hole.
Based on the Ruppeiner thermodynamic geometry, we calculate the thermodynamic scalar curvature $R$ for a hairy black hole of Einstein's theory conformally coupled to a scalar field in five dimensions without and with charges, respectively. By analogy with the behaviors of the usual ideal gas, Fermi gas, and Bose gas, we explore phenomenologically the microscopic information of black holes.
Moreover, with the help of the heat capacity, we analyze the thermal stability and phase transition behaviors of the black holes. Our results can be summarized as follows.
\paragraph{Scenario without charges}
\begin{itemize}
  \item For $q>0$, the negative thermodynamic scalar curvature $R<0$ implies that an attractive interaction dominates in black holes. The heat capacity is divergent only once, which implies that the black hole undergoes one  second-order phase transition from a locally unstable state to a locally stable one.
  \item For $q<0$, the thermodynamic curvature $R$ may be positive, negative, or zero, indicating that the microscopic feature of this black hole is analogous to that of the ideal anyon gas. Furthermore, the $P-V$ critical phenomenon exists. Beyond this critical point, the heat capacity is positive, which means that the black hole is in a stable state. Below the critical point, the heat capacity can diverge at two points, indicating that there exist two phase transitions for the black holes that tend to be stable eventually. At the critical point, the two phase transitions merge into one and finally the black holes are stable.
  \item On the co-existence curve, the thermodynamic curvature is negative for the large black hole whose behaviors look much like that of the ideal Bose gas, and the extremal large black hole resembles the ideal gas. However, the small black hole is similar to the ideal anyon gas, and the extremal small black hole is close to the ideal Fermi gas.
  \item Without equivalent AdS background, the thermodynamic scalar curvature is positive, resulting that the microscopic structure of this black hole matches that of the ideal Fermi gas. The heat capacity is negative for $q>0$, which means that the black hole is not stable. For $q<0$, the black hole undergoes one second-order phase transition and leads eventually to be more and more unstable.
\end{itemize}
\paragraph{Scenario with charges}
\begin{itemize}
  \item Regardless of $q>0$ or $q<0$, this situation is very similar to the case without charges under $q<0$. The sign of the thermodynamic curvature is positive, negative, or zero as expected, indicating that the microscopic feature of this black hole resembles that of the ideal anyon gas. Moreover, the black hole undergoes two phase transitions at least and ultimately it goes to a stable state.
\end{itemize}
In addition, the phrase -- intermolecular interaction -- is only a virtual description of black holes when the black holes are treated as a thermodynamic system. Completely from the thermodynamic point of view, we propose a new attempt to explore constituents
of black holes according to the type of interaction. Hence, we summarize that the microscopic structure for a hairy black hole of Einstein's theory conformally coupled to a scalar field in five dimensions is most likely analogous to that of the ideal Fermi, Bose or  anyon gas, which depends on the parameters $R$ and $q$. Moreover, we give the most probable explanation for the small-large black hole phase transition, i.e. it is the nature of the repulsive intermolecular interaction on the small black hole side that causes a very strong outward degenerate pressure for this hairy black hole in five dimensions. Finally, we should emphasize that our results are only preliminary and phenomenological description of the microstructure of black holes.

\section*{Acknowledgments}
This work was supported in part by the National Natural Science Foundation of China under grant No.11675081. The authors would like to thank the anonymous referee for the helpful comments that improve this work greatly.

\newpage

\section*{Appendix}
Substituting eqs.~(\ref{enth}),~(\ref{temp}),~(\ref{entr}), and~(\ref{pre}) into eq.~(\ref{rg}), we can directly calculate the metric components with coordinates $X^{\alpha}=(S,P)$ for a fixed $q$ as follows,
\begin{align}
g_{_{SS}}&=\frac{40 \sqrt[3]{2} \pi ^{5/3} P q \left(5 \pi ^2 q+4 S\right)^{2/3}+8 S \left(4 \sqrt[3]{\frac{2}{\pi }} P \left(5 \pi ^2 q+4 S\right)^{2/3}-3\right)-54 \pi ^2 q}{3 \left(5 \pi ^2 q+4 S\right) \left(8 \sqrt[3]{\frac{2}{\pi }} P S \left(5 \pi ^2 q+4 S\right)^{2/3}+10 \sqrt[3]{2} \pi ^{5/3} P q \left(5 \pi ^2 q+4 S\right)^{2/3}+9 \pi ^2 q+6 S\right)},\tag{A1}\\
g_{_{SP}}&=g_{_{PS}}=\left[{P+\frac{3 \sqrt[3]{\frac{\pi }{2}} \left(3 \pi ^2 q+2 S\right)}{2 \left(5 \pi ^2 q+4 S\right)^{5/3}}}\right]^{-1}, \tag{A2}\\
g_{_{PP}}&=0. \tag{A3}
\end{align}
Applying eqs.~(\ref{rgc}),~(\ref{rgr}), and~(\ref{rgs}) to our model, we work out the scalar curvature,
\begin{align}
&R=-\frac{\left(20 \pi ^2 P q \left(5 \pi ^2 q+4 S\right)^{2/3}+16 P S \left(5 \pi ^2 q+4 S\right)^{2/3}+9\times 2^{2/3} \pi ^{7/3} q+6\times2^{2/3} \sqrt[3]{\pi } S\right)}{\pi ^{2/3} \left(5 \pi ^2 q+4 S\right)^{4/3}} \nonumber\\
&\times \frac{\sqrt[3]{2} \left(15 \pi ^2 q+8 S\right) \left(2 \sqrt[3]{2} P \left(5 \pi ^2 q+4 S\right)^2+9 \pi ^{7/3} q \sqrt[3]{5 \pi ^2 q+4 S}+6 S \sqrt[3]{5 \pi ^3 q+4 \pi  S}\right)}{\left(8 \sqrt[3]{\frac{2}{\pi }} P S \left(5 \pi ^2 q+4 S\right)^{2/3}+10 \sqrt[3]{2} \pi ^{5/3} P q \left(5 \pi ^2 q+4 S\right)^{2/3}+9 \pi ^2 q+6 S\right)^3}. \tag{A4}
\end{align}
For simplification, we insert eqs.~(\ref{entr}) and~(\ref{pre}) into the above equation and finally obtain the thermodynamic scalar curvature in terms of the horizon radius $r_h$ and the effective AdS curvature radius $l$,
\begin{align}
R=-\frac{2l^2(4r_h^3+5q)}{3\pi^2 r_h^3(2r_h^3 l^2+q l^2+4r_h^5)}. \tag{A5}
\end{align}

\end{document}